\documentclass[prb,aps,epsfig,showpacs,superscriptaddress,twocoloumn,15point,reprint]{revtex4-1}
\usepackage{graphicx}
\usepackage{dcolumn}
\usepackage{bm}
\makeatletter
\usepackage{amsmath}
\usepackage{epstopdf}
\usepackage{soul,xcolor}
\usepackage{lineno,hyperref}
\usepackage{array}

\newcommand{\Rmnum}[1]{\expandafter\@slowromancap\romannumeral #1@}
\makeatother
\begin{document}

\title{\bf Fermi surface nesting driven anomalous Hall effect in magnetically frustrated Mn$_2$PdIn}

\author{Afsar Ahmed}\email{afsar.ahmed@saha.ac.in}
\author{Arnab Bhattacharya }
\affiliation{CMP Division, Saha Institute of Nuclear Physics, A CI of Homi Bhabha National Institute, Kolkata, WB-700064, India}
\author{Prashant Singh}
\affiliation{Ames National Laboratory, U.S. Department of Energy, Iowa State University, Ames, IA 50011, USA}
\author{Ajay Kumar}
\affiliation{Ames National Laboratory, U.S. Department of Energy, Iowa State University, Ames, IA 50011, USA}
\author{Tukai Singha}
\affiliation{CMP Division, Saha Institute of Nuclear Physics, A CI of Homi Bhabha National Institute, Kolkata, WB 700064, India}
\author{Anis Biswas} \email{anis@ameslab.gov}
\affiliation{Ames National Laboratory, U.S. Department of Energy, Iowa State University, Ames, IA 50011, USA}
\author{Yaroslav Mudryk}
\affiliation{Ames National Laboratory, U.S. Department of Energy, Iowa State University, Ames, IA 50011, USA}
\author{ Indranil Das} \email{indranil.das@saha.ac.in}
\affiliation{CMP Division, Saha Institute of Nuclear Physics, A CI of Homi Bhabha National Institute, Kolkata, WB 700064, India}

\begin{abstract}
Noncollinear magnets with near-zero net magnetization and nontrivial bulk electronic topology hold significant promise for spintronic applications, though their scarcity necessitates purposeful design strategies. In this work, we report a topologically nontrivial electronic structure in metallic Mn$_2$PdIn, which crystallizes in the inverse Heusler structure and exhibits a spin-glassy ground state with quenched magnetization. The system features Weyl-type band crossings near the Fermi level and reveals a novel interplay among momentum-space nesting, orbital hybridization, and spin-orbit coupling. Comprehensive transport measurements uncover a pronounced anomalous Hall effect (AHE) in Mn$_2$PdIn. The observed quadratic relationship between the longitudinal and anomalous Hall resistivities highlights the intrinsic Berry curvature contribution to AHE. These findings establish inverse Heusler alloys as compelling platforms for realizing noncollinear magnets that host Weyl-type semimetallic or metallic phases-combining suppressed magnetization with robust electronic transport-thereby offering a promising route toward their seamless integration into next-generation spintronic devices.
\end{abstract}

\maketitle

Noncollinear magnetic ordering such as skyrmions\cite{fert2017magnetic,tokura2020magnetic,fert2013skyrmions,gallagher2017robust,ozawa2017zero}, merons\cite{puphal2020topological,yu2018transformation}, hedgehogs\cite{fujishiro2019topological,zou2020topological} and spin spirals\cite{yang2021chiral,park2023tetrahedral,ghimire2018large}, challenge the conventional dichotomy of parallel or anti-parallel spin arrangements in traditional ferromagnets and antiferromagnets. Such non-collinear spin arrangements can exhibit ferroic responses, like the spontaneous Hall effects, without significant net magnetization while mitigating the stray field issues typically associated with ferromagnets \cite{suzuki2016large,shekhar2018anomalous,ghimire2018large}. Notably, these magnets, including Weyl semimetal Mn$_3$Sn\cite{nakatsuji2015large,li2023field}, exhibit anomaly in transverse resistivity ($\rho_{xy}$) originating in the nontrivial electronic states proximate to the Fermi level, where large intrinsic Berry curvature drives a substantial AHE. However, the fundamental understanding of such non-collinear magnets exhibiting near-zero net magnetic moment yet delivering a robust transport response driven by the nontrivial electronic topology remains in its infancy\cite{vsmejkal2017route}. This gap largely stems from the scarcity of materials that combine noncollinear magnetic ordering with Weyl semimetallic/metallic (WSM) electronic states \cite{puphal2020topological,Ahmed2025,bhattacharya2024spin}, underscoring the need to explore novel materials fostering this synergy with robust electrical transport signal of magnetism for seamless interfacing with electronic devices.

The emergence of Weyl fermions necessitates the breaking of either inversion ($\mathcal{P}$) or time-reversal ($\mathcal{T}$) symmetry and can lead to  large intrinsic AHE.   Building on these prerequisites, Mn-based inverse Heusler alloys (iHA), featuring inherently broken $\mathcal{P}$ and multiple inequivalent magnetic sublattices \cite{graf2011simple}, have emerged as a compelling platform for concerned synergy.  The antiferromagnetic exchange interaction between the tetrahedrally coordinated Mn atoms and their highly localized octahedral counterparts stabilizes a ferrimagnetic ordering with a small net magnetization \cite{shi2018prediction,bhattacharya2024spin,PhysRevB.83.174448,PhysRevB.95.060410}. Despite vanishing magnetization, the intrinsic magnetism mediated broken-$\mathcal{T}$ lifts the spin degeneracy of electronic bands, facilitating the formation of Weyl points. Meanwhile, the moderately broken $\mathcal{P}$, while preserving the topological nature of the Weyl points, shifts them in three-dimensional momentum space to different energy levels from the Fermi level, precluding the emergence of a nodal-line semimetal phase\cite{PhysRevB.85.165110}. Notably, the 4$d$-transition-element-based iHA holds distinct advantages of reduced anti-site disorder and enhanced spin-orbit coupling (SOC) effects over their 3$d$ counterparts.

In this context, we present Mn$_2$PdIn, a ternary derivative of cubic spin-glassy alloy Mn$_3$In\cite{chatterjee2020glassy} with peculiar electronic band structure featuring clear Fermi surface nesting exhibits magnetically frustrated ground state and significant AHE through combined experimental and theoretical study. Employing systematic magnetometry investigation, we reveal a stabilization of spin cluster-glassy ground state with a spontaneous magnetization of 0.46 $\mu_B$/f.u., following a ferrimagnetic/ferromagnetic-like ordering near room temperature.  These results underscore Mn-based iHA as a unique platform for exploring the interplay of noncollinear magnetism and topological electronic states.

\begin{figure}[h]
\begin{center}
\includegraphics[width=.4\textwidth]{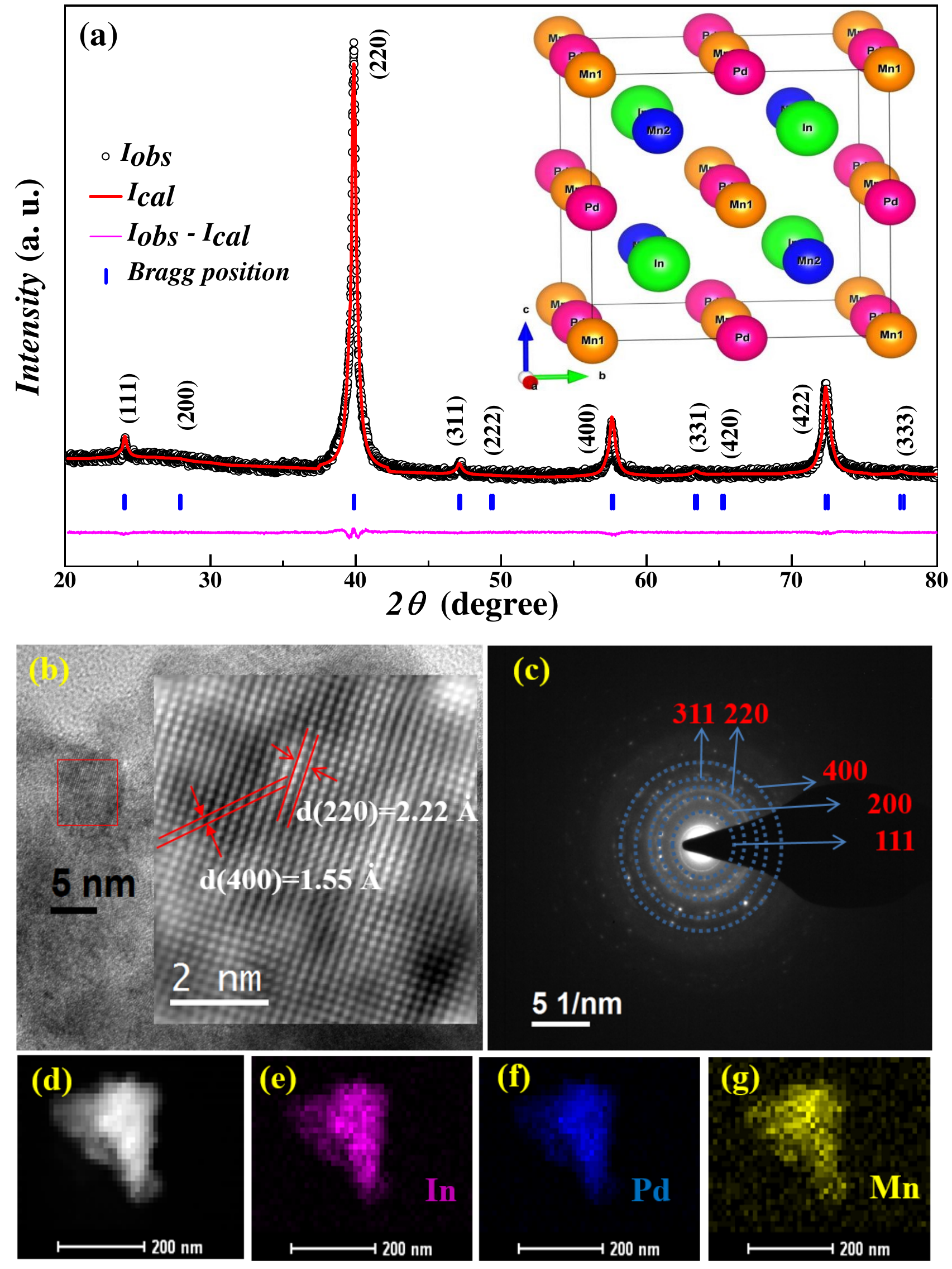}
\end{center}
\caption[]
{\label{SF2} (a) Rietveld refinement of room temperature XRD pattern of Mn$_2$PdIn. Inset shows crystal structure of the Mn$_2$PdIn compound. (b) High resolution TEM image, inset shows Fourier filtered image. (c) Selected-area electron diffraction (SAED) pattern. (d) STEM-HAADF image of the particle on which the elemental mapping was taken; elemental profile of (e) In (f) Pd and (g) Mn.}
\label{F4}
\end{figure}
 Polycrystalline samples of Mn$_2$PdIn were synthesized through a conventional arc-melting process using high-purity (99.9$\%$) constituent elements in an argon (Ar) atmosphere. Following the initial synthesis, the samples underwent annealing at 800$^\circ$ C for one week within evacuated sealed quartz tubes to ensure better homogeneity. The crystallographic structure and phase purity of the samples were evaluated using X-ray diffraction (XRD) at room temperature, employing Cu-K$_\alpha$ radiation with a Rigaku TTRX-III diffractometer. To verify the elemental composition and homogeneity of the samples, transmission electron microscopy (TEM) equipped with an energy-dispersive X-ray (EDX) spectrometer was utilized. The DC magnetic properties were investigated using a superconducting quantum interference device (SQUID-VSM) across a temperature range of 2 to 380 K and a magnetic field range of $\pm$ 7 T. The ac magnetic susceptibility measurements were conducted using an ac-equipped SQUID to provide further insight into the magnetic behavior of the samples. Magnetotransport properties were measured on samples with approximate dimensions of 4.50$-$4.60 mm (length), 2.00$-$2.05 mm (width), and 0.30$-$0.35 mm (thickness). These measurements were carried out using a 9T physical property measurement system (PPMS) in a standard four-probe arrangement, applying a DC current of 20 mA. Prior to the measurements, the ohmic nature of the current and voltage probe contacts, established with gold (Au) wires, was confirmed across the entire temperature range. For the assessment of field dependence of the resistivity at each temperature, the samples were cooled from room temperature to the desired measurement temperature in the absence of an applied magnetic field (H). The transverse resistivity ($\rho_{xy}$) was symmetrized using the relation, $\rho_{xy} = \left[\rho_{xy}(+H) - \rho_{xy}(-H)\right]/2$.


\begin{table}[ht]
\centering
\caption[]
{\label{ST1} Table of crystallographic parameters, obtained from full Rietveld refinement analysis of room temperature XRD.}
   \begin{tabular}{p{7.5 cm}}
   \hline
   \hline
   Lattice parameters: a = b = c = 6.399(5) $\AA$\\
   Space group: F$\overline{4}$3m (No. 216)\\
   $R_f$ = 5.12, $R_{Bragg}$ = 8.03, $\chi^2$ = 1.67 \\
   \noalign{\smallskip}
   \end{tabular}
\begin{tabular}{p{2.0 cm}p{4.0 cm}p{1.0 cm}}
&   \textbf{Atomic Coordinates}&\\
\end{tabular}
\begin{tabular}{p{1.5 cm} p{1.5 cm} p{1.5 cm} p{1.5 cm} p{1.5 cm}}
\hline
\noalign{\smallskip}
Atom  & Wyckoff position & x & y & z \\
\noalign{\smallskip}
\hline
\noalign{\smallskip}
Mn$_{II}$ & 4c & 0.25 &  0.25 &  0.25 \\
Pd & 4d & 0.75 &  0.75 &  0.75 \\
Mn$_{I}$ & 4b & 0.50 & 0.50 &  0.50  \\
In & 4a &  0.00 & 0.00 &  0.00 \\
\noalign{\smallskip}
\hline
\hline
\end{tabular}
\end{table}

Figure \ref{SF2} (a) presents the Rietveld refinement of the room-temperature XRD pattern for Mn$_2$PdIn, obtained using a Rigaku TTRX-III diffractometer. The refinement confirms the inverse-Heusler alloy (iHA) structure [inset of Fig. \ref{SF2} (a)] belonging to the cubic $F\Bar{4}3m$ space group (216) with a lattice parameters $a$ = 6.399(5) $\AA$, consistent with prior report\cite{xu2016magnetic}. The fitting parameters obtained from Rietveld refinement analysis are enlisted in Table \ref{ST1}. This structure adheres to the stoichiometric arrangement of $XYX'Z$, with $X$, $X'$, $Y$, and $Z$ occupying 4$c$, 4$d$, 4$b$ and 4$a$ Wyckoff positions over four interpenetrating FCC sublattices. Notably, the absence of the (200) superlattice peak validates the iHA structure, consistent with the relative electronegativity of Mn and Pd \cite{graf2011simple}.  Figure \ref{SF2} (b) showcase high resolution TEM (HRTEM) image of Mn$_2$PdIn. The inset of Fig. \ref{SF2}(b) highlights the Fourier-filtered image of the specified region shown by the red box in the HRTEM image [Fig. \ref{SF2}(b)]. The interplanar spacings of 2.22 $\AA$ and 1.55 $\AA$ were identified as the [220] and [400] lattice planes, respectively. The polycrystalline nature of the compound is evident from the ring-shaped selected area electron diffraction (SAED) pattern, as shown in Fig. \ref{SF2}(c). The scanning TEM high-angle annular dark field (STEM-HAADF) image of the particle is presented in Fig. \ref{SF2}(d), on which elemental mapping and EDX analysis were performed. The elemental mapping of In, Pd, and Mn confirms the homogeneity of the compound, as shown in Fig. \ref{SF2}(e)-(g). Through XRD study, no impurity phase was detected which was further supported TEM-EDX elemental mapping.

\begin{figure}[t]
\begin{center}
\includegraphics[width=.49\textwidth]{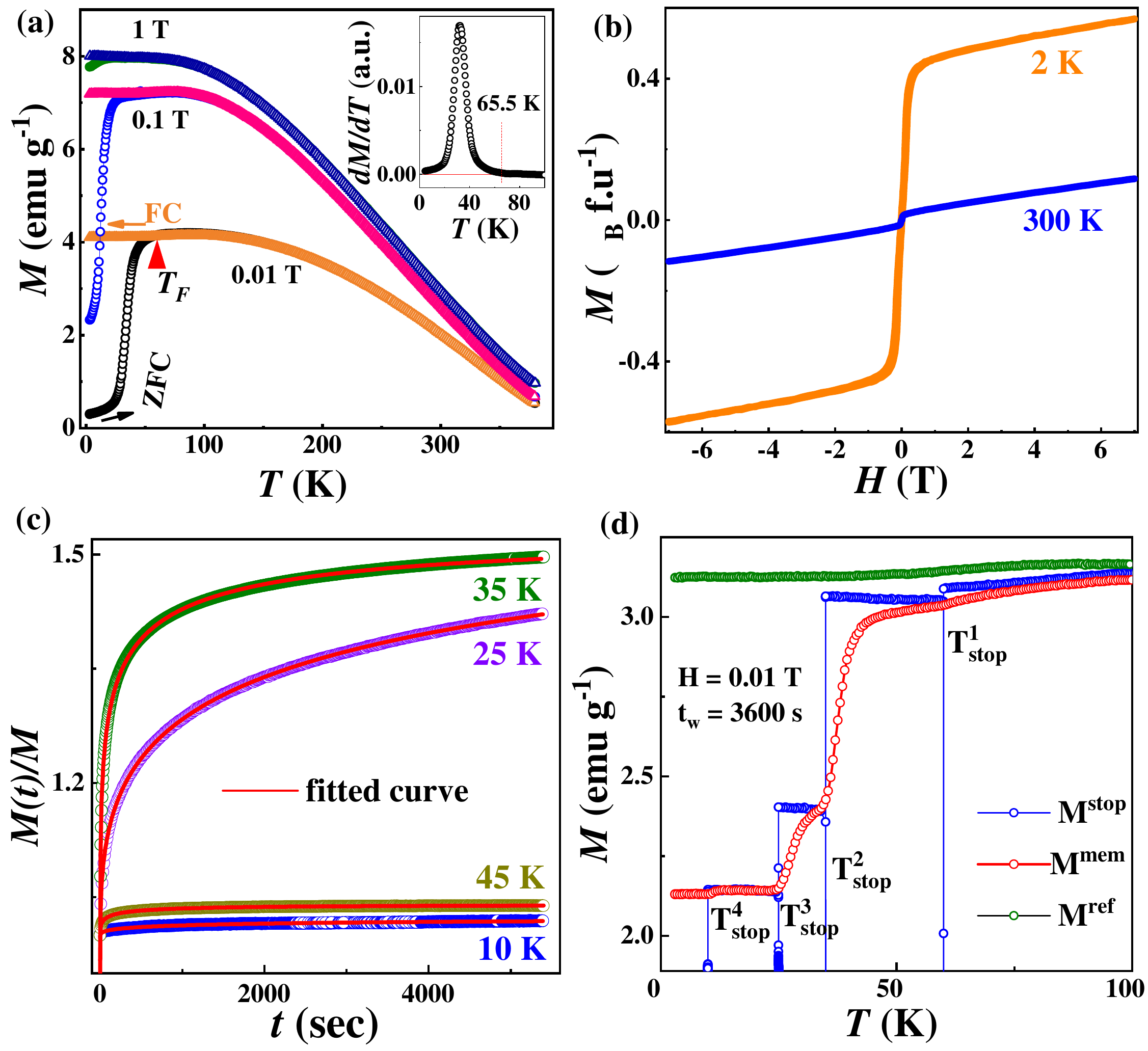}
\end{center}
\caption[]
{\label{f1} (a) $M(T)$ curves under ZFC and FC protocols for various applied fields. Inset shows the derivative of 0.01 T ZFC $M(T)$ curve. (b) $M(H)$ curves at 2 and 300 K. (c) Time dependence of $M$ for different $T$ under $H$ = 0.01 T. (d) Field-cooled-field stop memory curves.}
\label{F4}
\end{figure}

Figure \ref{f1}(a) depicts the DC-thermomagnetic curves, $M(T)$, measured under different applied fields ($H$) using zero-field-cooled (ZFC) and field-cooled (FC) protocols, showing an increase in $M$ around room temperature, characteristics of a paramagnetic (PM) to ferromagnetic (FM)/ferrimagnetic (FiM)-like state, as reported earlier \cite{xu2016magnetic}.  Notably, with lowering of temperature, a distinct bifurcation between $M_{ZFC}$ and $M_{FC}$ appears below the freezing temperature ($T_{F}$) $\sim$ 65.5 K. Remarkably, the bifurcation is persistent under moderately high $H$ of 0.1 T while $T_{F}$ shifts towards lower temperature with increasing $H$, a behaviour typical for glassy phase \cite{samanta2018reentrant,PhysRevB.99.174410,chatterjee2020glassy,PhysRevLett.110.127204}.

The isothermal magnetization $M(H)$ at 2 K exhibits a soft FM/FiM-like behavior with negligible hysteresis, while the small spontaneous $M(H)$ at 300 K near $H$ = 0 indicates a persistent FM/FiM contribution, consistent with earlier report\cite{xu2016magnetic} [Fig. \ref{f1}(b)]. Even under a 7 T magnetic field at 2 K, the magnetization does not reach saturation but instead continues to exhibit field dependence, suggesting a tendency toward FiM behavior. However, it is worth mentioning that the derived  saturation moment of 0.46 $\mu_B$/f.u from the linear fitting of $M(H)$ curve in the 5 - 7 T range at 2 K, which is significantly lower than the expected value of 3 $\mu_B$/f.u based on the Slatter-Puling rule ($M_{sat}$ = $N_v-$24; $N_v = 27$ is the total number of valence electron number), for a collinear FM ordering \cite{skaftouros2013generalized}. This deviation can be attributed to  the presence of intersublattice antiferromagnetic correlation between two different Mn atoms located at the octagonally coordinated 4$b$ site in the Mn-In sublattice and the tetragonal 4$c$ site in the Mn-Pd plane, leading to a net FiM interaction\cite{bhattacharya2024spin,p35,p110,PhysRevB.83.174448}. The presence of competing interactions can lead to magnetic frustration and glassy magnetic state.

To get insights into the possible magnetic glassy state, we performed magnetic relaxation measurements at various temperatures under $H$ of 10 mT, following ZFC from above $T_F$ \cite{PhysRevLett.65.2674,PhysRevB.99.174410,PhysRevB.94.104414}. The time-dependent magnetization $M(t)$, shown in Fig. \ref{f1}(c), reveals a gradual increase over time, a phenomenon characteristic of the magnetic aftereffect\cite{PhysRevB.94.104414,PhysRevB.99.174410,phillips1996stretched,PhysRevB.94.104414,PhysRevLett.72.3270,mydosh1993spin}. The curves are well-described by Kohlrausch-Williams-Watts equation: $M(t) = M_0(1+a exp[-(t/\tau)^\beta])$, where $M_0$ is intrinsic magnetization, $\tau$ is characteristics relaxation time and $\beta$ is shape parameter accounting for the energy barrier involved in the relaxation process. The extracted values of $\tau$ and $\beta$, listed in Table \ref{Relax}, show that $\beta$ remains in the range of 0.30 to 0.34 for all temperatures, consistent with the literature on SG systems \cite{PhysRevB.99.174410,phillips1996stretched,PhysRevB.94.104414,PhysRevLett.72.3270,mydosh1993spin}.
\begin{table}[t]
	\begin{center}
		\caption{Fiting parameters $t$, $\beta$, $\tau$ for different temperatures using the Kohlrausch-Williams-Watts equation.
        \label{Relax}}
		\begin{tabular}{p{3 cm}p{2.5 cm}p{1.5 cm}p{1 cm}}
			\hline
			\hline
			\noalign{\smallskip}
			Temperature (K) \indent & $\beta$ \indent & \indent $\tau$ (s)  \\
			\noalign{\smallskip}
			\noalign{\smallskip}
			\hline
			\noalign{\smallskip}
			10 & 0.30(2) & 2284 \\
			25 & 0.34(8)& 3339 \\
			35 & 0.32(1) & 229 \\
		    45 & 0.33(4) & 91\\
			\noalign{\smallskip}
			\noalign{\smallskip}
			\hline
			\hline
		\end{tabular}
	\end{center}
\end{table}
Alongside magnetic relaxation, another notable aspect of the spin-glassy state is the manifestation of magnetic memory effects\cite{chatterjee2020glassy,PhysRevB.94.104414}. The FC memory effect was measured according to Sun \textit{et al.}, where the sample is initially cooled down from room temperature to 2 K under $H$ of 10 mT with intermediate temperature halts at $T_{stop}$ (= 65, 35, 25 and 10 K) $<T_f$ for 1.5 hr while the field was switched off during each halting period. Following the lapse of $t_w$, the field was switched back on and the cooling was resumed while recording the temperature dependence of magnetization $M^{\mathrm{stop}}$ [Fig. \ref{f1}(d)]. Following the cooling up to 2 K, the sample was heated to room temperature without any halts while recording $M^{\mathrm{mem}}$ and another conventional FC magnetization data was recorded as a reference ($M^{ref}$). The $M^{\mathrm{mem}}$ tries to retrace $M^{\mathrm{stop}}$ exhibiting anomalous bending at all $T_{stop}$, suggesting that the system remembers it previous history. This behaviour is consistent and resonates well with the literature of SG systems.

\begin{figure}[t]
\begin{center}
\includegraphics[width=.49\textwidth]{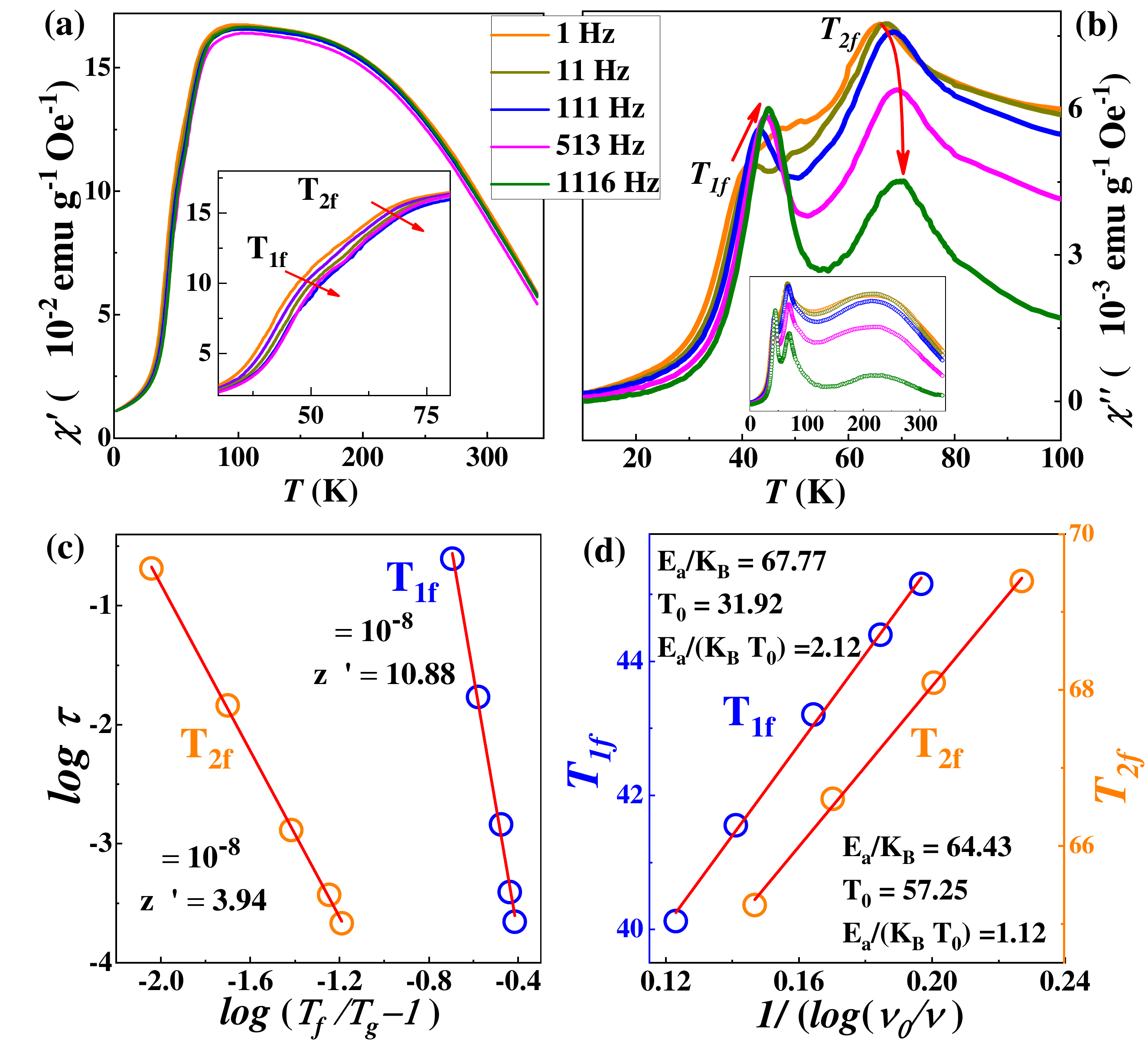}
\end{center}
\caption[]
{\label{f2}(a),(b) Temperature variation of the real ($\chi'(T)$) and imaginary ($\chi^{\prime \prime}(T)$) components of ac-susceptibility for different $f$. Insets: (a) and (b) show an enlarged view of $\chi'(T)$ at low $T$ and $\chi^{\prime \prime}(T)$ over 2 - 350 K, respectively. (c) $f$ dependence of both freezing temperatures $T_{f1}$ and $T_{f2}$ where log$\tau$ is plotted with log$(T_f/T_g-1)$. Red lines are the fitted curve using power law. (d) Frequency dependence for both the freezing temperatures plotted as $T_{1f}$ and $T_{2f}$ vs. 1/log$(\nu_0/\nu)$. Red lines are fitted curves.}
\end{figure}

Following the DC-thermomagnetic observations, we investigated the $\chi_{ac}(T)$ under an excitation ac-field of 6 Oe for a broad frequency ($f$) range. Figure \ref{f2}(a),(b) depicts the real ($\chi'$) and imaginary ($\chi''$) part of $\chi_{ac}(T)$, respectively, for $f$ ranging over three orders of magnitude. The $\chi'(T)$ showcase two distinct $f$-dependent positive peak-shift behavior at $T_{1f}$ and $T_{2f}$ [inset of Fig. \ref{f2}(a)]. While $T_{1f}$ does not manifest in DC measurements, $T_{2f}$ resonates with $T_{F}$. Notably, $\chi''(T)$ exhibits frequency-dependent peaks at $T_{1f}$ and $T_{2f}$, along with an $f$-independent peak around 250 K, indicating dissipative losses associated with successive SG transitions and domain formation, possibly following long-range magnetic ordering \cite{p112,p113,p114}. The anomaly at $T_{1f}$ in $\chi'(T)$, which is absent in DC magnetometry, suggests the presence of nanoferromagnetic clusters. These clusters likely become suppressed or dissolve under the application of a small DC field (10 mT), possibly due to field-induced alignment of magnetic moments or the reduction of spin fluctuations. For a conventional spin-glassy system, the relative shift of freezing temperature $T_f$ per decade of $f$ is expressed as \cite{mydosh1993spin}, $\delta T_f$ = $\Delta T_f/(T_f\Delta log f)$. In the present case, the estimated value of $\delta T_{1f}$ = 0.054 and $\delta T_{2f} = 0.017$ lies typically in spin cluster-glass limit for both transitions\cite{mydosh1993spin}.
For SG systems, the relaxation time $\tau (=f^{-1})$ at a particular $T_f$ is related to the spin-spin correlation length $\xi$ by a power-law divergence relation, $\tau = \tau_0 \left(\frac{T_f-Tg}{T_g}\right)^{-z \nu'}$, where $\tau_0$ and $T_g$ are the single spin-flip relaxation time and SG transition temperature for $f$ = 0, respectively\cite{mydosh1993spin,RevModPhys.49.435}. While $z \nu'$ is the dynamic critical exponent and  $\nu'$ is the critical exponent of correlation length, $\xi = (T_f/T_g-1)^{-\nu'}$. Typically, the $z \nu'$ lies between 4 to 12 for SG systems\cite{PhysRevB.68.214422,PhysRevB.86.064412}. In Fig. \ref{f2}(c), from our experimental results we obtain $z \nu' \simeq$ 11 and $\tau_0 \simeq$ 10$^{-8}$ for $T_{1f}$ and $z \nu' \simeq$ 4 and $\tau_0 \simeq$ 10$^{-8}$ for $T_{2f}$, respectively. For both cases, $\tau_0$ is a few orders higher than the value for canonical spin-glass ($\sim 10^{-12}-10^{-13}$), suggesting a typical spin cluster-glass nature for the system\cite{PhysRevB.68.214422}. To further verify the nature of the SG state, we analyzed the frequency dependence of $T_f$ employing the Vogel-Fulcher relation, $\nu = \nu_0  exp\left(\frac{-E_a}{K_B(T_f-T_0)}\right)$, where $\nu_0$, $E_a$, and $T_0$ are characteristic attempt frequency, activation energy and Vogel-Fulcher temperature, respectively \cite{mydosh1993spin,RevModPhys.49.435}. The linearity relation of 1/log($\nu_0$/$\nu)$ vs. $T_f$ plot, depicted in Fig. \ref{f2}(d), yields $E_a/K_BT_0 \simeq 2.12$ and 1.12 for $T_{1f}$ and $T_{2f}$, respectively, affirming the cluster spin-glassy nature for Mn$_2$PdIn \cite{mydosh1993spin}.

To understand ground state magnetic properties, we performed first principle  electronic structure calculations for the samples . In this context, we employed density functional theory (DFT) method as implemented with Vienna Ab-initio Simulation Package (VASP)  where the valence interaction among electrons was described by a projector augmented-wave (PAW) method \cite{ref1, ref2} with an energy cutoff of 520 eV for the plane-wave orbitals. We used a $7 \times 7 \times 7$ ($11 \times 11 \times 11$) Monkhorst-Pack $k-$mesh for Brillouin zone sampling of cubic unit-cell during structural optimization (electronic-relaxation) \cite{ref3}. A very high accuracy of total energy and force convergence, i.e., $10^{-8}$ eV/cell and $10^{-6}$ eV/\AA, was set to get fully relaxed ionic-positions and electronic charges. We employ the Perdew-Burke-Ernzerhof (PBE) exchange-correlation functional in the generalized gradient approximation (GGA) \cite{ref4}. In (semi)local functionals, such as GGA, the f-electrons are always delocalized due to their large self-interaction error \cite{ref5, ref6}.

Our calculations reveal that the ferri-magnetic (FiM) configuration is energetically preferred over the anti-ferromagnetic (AFM) or ferromagnetic (FM) state in Mn$_2$PdIn with a formation enthalpy of -67.95 meV/atom, resulting in a net magnetic moment of 0.39 $\mu_B$ per formula unit, which agrees quite well with experimentally obtained value from $M(H)$ curve. This arises from the opposing magnetic moments of 3.75 $\mu_B$ (Mn1) and -3.48 $\mu_B$ (Mn2), with small induced magnetic contributions from Pd (0.057 $\mu_B$) and In (0.018 $\mu_B$). Based on competing energetic FiM and FM, a mean-field approach which emphasizes spin-spin pair correlations is used to calculate the freezing temperature $T_F$ of metallic spin glasses \cite{ref7, ref8, ref9, ref10}. We used the relationship: \[T_F = \frac{1}{k_B N} \sum_{i,j} \sqrt{J_{i,j,\text{avg}}^2}\] to approximate the spin-glass temperature, where $J_{i,j,\text{avg}}^2$ is the average effective indirect exchange coupling over all possible configurations of magnetic ions and nonmagnetic ions, $N$ is the total number of spins, and $k_B$ is Boltzmann's constant \cite{ref7}. The resulting estimate using DFT calculated $J[\text{FiM-AFM}] = -0.04295$ eV, eight spin centers ($N=8$) in the unit-cell, and $k_B$ (8.617 $\times 10^{-5}$ eV$\cdot$K$^{-1}$) was found to be 62.3 K, which is in good agreement with the experimentally observed $T_F = 65.5$ K [see Fig. \ref{F4}(a)].

\begin{figure*}[t]
\begin{center}
\includegraphics[width=.95\textwidth]{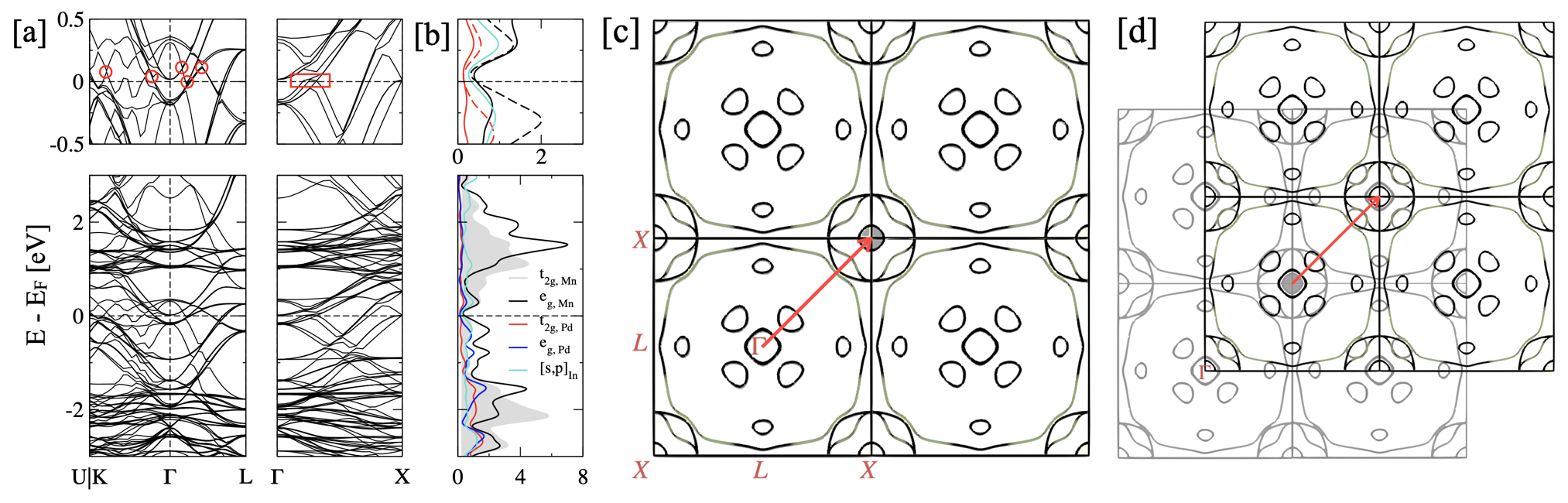}
\end{center}
\caption{(a, b) Electronic band structure and atom- and orbital-resolved density of states (DOS) for Mn$_2$PdIn including spin-orbit coupling (SOC). The band structure reveals multiple band crossings near the Fermi level, with an electron-like Mn $e_g$ band at $\Gamma$ and flatter, hole-like bands near $L$ and $X$ exhibiting significant Pd $t_{2g}$ and Mn $t_{2g}$ character, creating quasi-nested Fermi surface segments. The projected DOS confirms dominant Mn $e_g$ contributions at the Fermi level, alongside hybridization with Pd $d$- and In $\{s,p\}$-states, which enhances inter-orbital scattering pathways. Fermi surface contours of Mn$_2$PdIn with SOC illustrating nesting-driven interband scattering, where (c) the red arrow highlights a prominent nesting vector connecting electron-like pockets at the $\Gamma$ point with hole-like pockets near the $X$ point in the $k_{x}-k_{y}$ plane, and (d) superimposed Brillouin zones emphasize the commensurate nature of the $\Gamma$-$X$ nesting arising from Mn $e_g$-derived conduction bands overlapping with flat, hybridized valence bands near $X$, thereby facilitating momentum-dependent scattering linked to the anomalous Hall effect.}
\label{P3}
\end{figure*}

The band structure in Fig. \ref{P3}(a) shows multiple moderately dispersive bands crossing the Fermi level ($E_{\text{Fermi}}$), particularly along high-symmetry lines such as $\Gamma \rightarrow X$. Around the $\Gamma$ point, an electron-like parabolic band sits just above $E_{\text{Fermi}}$, characterized primarily by Mn $e_{g}$ orbital weight. In contrast, near the $L$ point, a flatter hole-like band dips just below $E_{\text{Fermi}}$, with strong contributions from Mn $t_{2g}$ and Pd $d$-orbitals. This energy-momentum configuration establishes a nesting condition across the Brillouin zone via a vector $\bm{q}_{\text{nest}} \sim \Gamma \rightarrow X$, which connects parallel Fermi surface contours separated by approximately $0.5$ - $0.7$~\AA$^{-1}$ in reciprocal space. These features create quasi-parallel energy contours, enhancing the nesting susceptibility between the $\Gamma$ and $L$ points. Moreover, in systems like Mn$_2$PdIn, which break time-reversal symmetry due to magnetic ordering, such crossings are strong candidates for Weyl points, especially in the presence of spin-orbit coupling (SOC).
The SOC in this system, introduced primarily by the Pd atoms, acts to lift degeneracies and generate small energy gaps (on the order of 20$-$50 meV). The presence of these band crossings in close proximity to the Fermi level is crucial for stabilizing symmetry-protected Weyl nodes, which may serve as localized sources and sinks of Berry curvature, thereby contributing significant intrinsic AHC. The red-boxed region along the $\Gamma \rightarrow X$ and $\Gamma \rightarrow K$ points highlights another important feature: the presence of quasi-flat bands that span the Brillouin zone. These bands exhibit low group velocities, resulting in an enhanced density of states near the Fermi level. From a transport standpoint, flat bands in close energetic alignment with linearly dispersing Weyl-like bands can amplify the magnitude of Berry curvature due to enhanced dwell time of quasiparticles and increased interband coherence. The strong orbital selectivity of these bands, with dominant Mn $e_{g}$ and Pd $t_{2g}$ contributions as shown in Fig. \ref{P3}(b), facilitates large matrix elements for anomalous velocity transitions. Notably, in Fig. \ref{P3}(b), the PDOS confirms that the states at the Fermi level are dominated by Mn $e_{g}$ orbitals, with a sharp DOS peak indicating a high electronic density and associated susceptibility to magnetic and transport instabilities.

The flatter bands from $\Gamma \rightarrow X$ display increased hybridization from Pd $t_{2g}$ and $e_{g}$ states, whose symmetry and localization promote band anticrossings. In contributions are visible through weakly dispersive $\{s, p\}$-states that mediate broader hybridization across the valence manifold. The resulting orbital entanglement near $E_{\text{Fermi}}$ leads to multiple avoided crossings, especially along $\Gamma \rightarrow L$, as observed in the band structure- a hallmark of SOC-induced band distortion. The combined proximity of Weyl-like crossings and flat bands near E$_{\rm Fermi}$, together with orbital hybridization and spin-orbit splitting, may enable a large momentum-resolved asymmetry in electronic velocity. Quantitatively, the electronic structure of Mn$_2$PdIn supports the possibility of a robust intrinsic anomalous Hall effect.

 This electronic complexity culminates in the Fermi surface shown in Fig. \ref{P3}(c) \& (d), where nested electron- and hole-like pockets emerge around the $\Gamma$ and $X$ points, respectively. The red arrow traces a well-defined nesting vector $\bm{q}$, enabling low-energy electronic transitions between states of opposite curvature and group velocity. These transitions are allowed because of the energetic closeness (within $\pm$0.2 eV) and similar orbital symmetry, with the pockets forming nearly commensurate contours repeated across Brillouin zones due to the cubic symmetry of the system. The SOC, introduced via the heavy Pd and In atoms, further lifts degeneracies near band crossings, producing small but finite gaps where previously degenerate states are now separated by orbital mixing. This splitting introduces directional asymmetry in the velocity matrix elements $\langle u_{n\bm{k}} \vert v_x \vert u_{m\bm{k}} \rangle$, where $n$, $m$ index the nested bands. These asymmetries generate a momentum-resolved imbalance in carrier distribution under an external electric field, producing a finite transverse conductivity $\sigma_{xy}$. Importantly, the proximity of the Mn $e_g$-dominated conduction band to quasi-flat valence bands at $X$ not only amplifies the density of interband states but also increases the phase space for asymmetric scattering. The nesting vector $\bm{q}_{\text{nest}} \sim \Gamma \rightarrow X$ bridges bands with strong orbital contrast (Mn $e_g \leftrightarrow$ Pd $t_{2g}$), enabling scattering pathways that break inversion symmetry locally in momentum space.

We anticipate significant  AHE in Mn$_2$PdIn as a natural outcome of the combined effects of Fermi-surface nesting, SOC-induced band hybridization, and orbital anisotropy. The alignment of electron-like bands at $\Gamma$ with quasi-flat hole-like bands near $X$, their orbital composition (especially Mn $e_g$ and Pd $t_{2g}$), and the scattering permitted by these nesting vectors all drive momentum-dependent asymmetric conduction. To further confirm role of electronic-structure, we evaluated intrinsic AHE ($\sigma_{xy, \text{int}}^A$) of Mn$_{2}$PdIn using the Kubo formalism, as outlined by Yao \textit{et al.} and others \cite{Yao2004, Fuh2011,Huang2015}. The imaginary part of the off-diagonal elements of the optical conductivity is first computed, followed by the determination of the real part through a Kramers-Kronig transformation. This approach, based on linear response theory, relates the AHC to the electronic structure by integrating the velocity operators and wavefunctions over the Brillouin zone. The intrinsic AHC corresponds to the static limit of the off-diagonal optical conductivity, i.e., $\sigma_{xy, \text{int}}^1 (\omega=0)$. Alternatively, AHC can also be obtained by integrating the Berry curvature over the Brillouin zone, an approach that has been shown to yield numerically equivalent results \cite{Guo2014}. \cite{Guo2014}. To ensure accuracy, we employed a dense $k$-point mesh for AHC calculations. Multiple fine $k$-point meshes were used, with the densest being $70 \times 70 \times 70$ in $F4\bar{3}m$ Brillouin zone. The AHC was fitted to a polynomial to obtain the converged theoretical value, i.e., the extrapolated value at dense $k-$mesh, following the approach in Ref. \cite{Fuh2011}.

\begin{figure*}
\begin{center}
\includegraphics[width=0.95\textwidth]{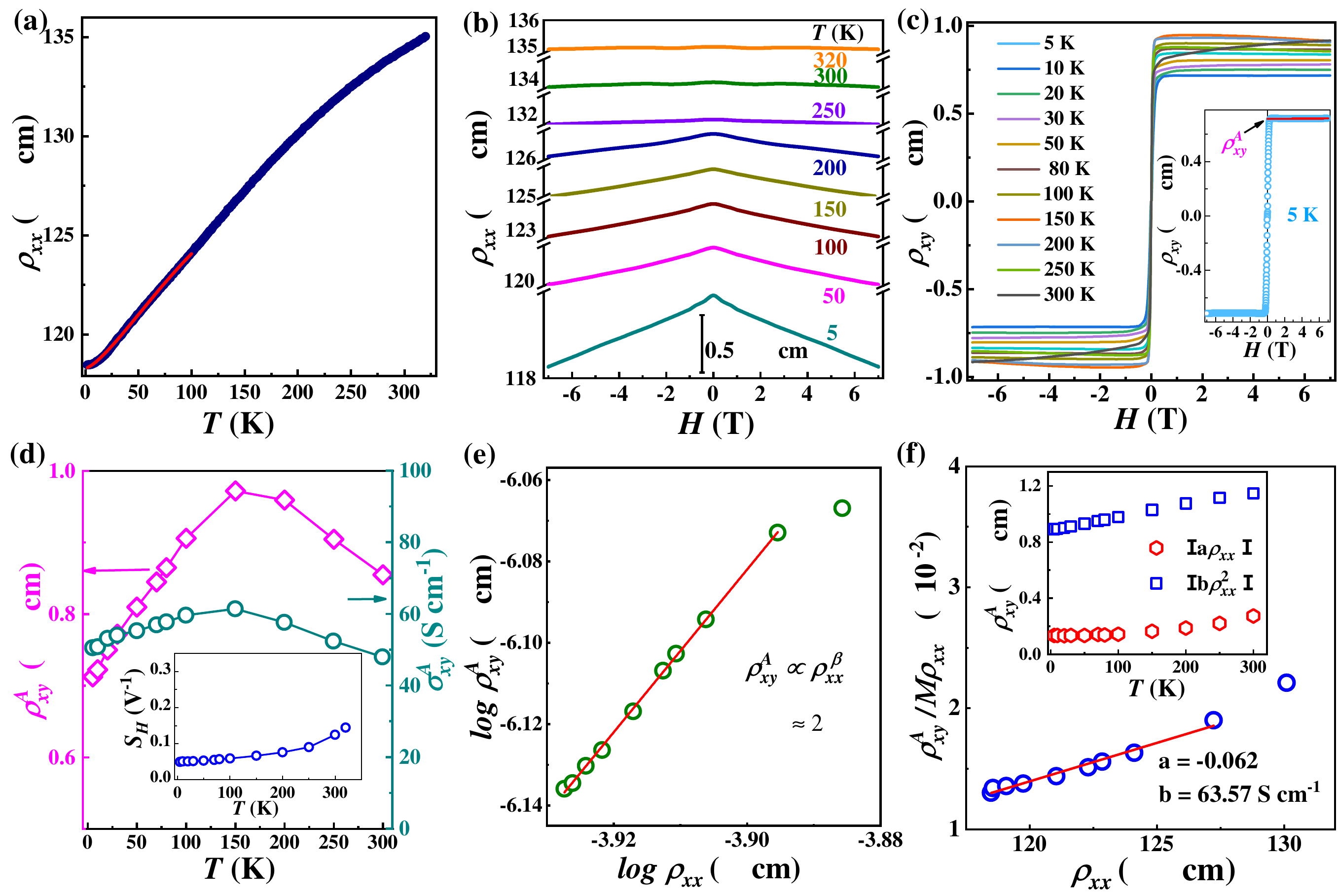}
\end{center}
\caption[]
{\label{F3}(a) $\rho_{xx}(T)$ in zero field. Red line is the fitted curve (b) $\rho_{xx}(H)$ isotherms for various $T$. (c) Transverse resistivity for different $T$.  Inset shows a linear fit of $\rho_{xy}$ in the high-field range ($H$ = 5 - 7 T), yielding $\rho_{xy}^A$. (d) Variation of $\rho_{xy}^A$ and $\sigma_{xy}^A$ with $T$. Insets show the  $T$ variation of scale factor $S_H$. (e) Power-law scaling of $\rho_{xy}^A$ with $\rho_{xx}$. (f) TYJ scaling. Inset shows the $T$ variation of $int$, $sj$ and skew scattering component of AHC.}
\label{F3}
\end{figure*}
Notably, electronic-structure in Fig. \ref{P3} reveals that bands near the Fermi level consist of highly dispersive Mn$-d$, Pd$-d$, and In$-p$ hybridized bands with very low electronic density. As a result, the intrinsic AHC, as obtained from DFT combined with Kubo formalism, is not very large, measuring approximately 132 S cm$^{-1}$. However, when the energy is reduced (increased) to -2.1 (0.8 eV), $\sigma_{xy,int}^A$ increases dramatically to about 937 S cm$^{-1}$ ($\sim$1003 S cm$^{-1}$). The enhanced AHC is primarily attributed to the increased electronic DOS in these energy regions, leading to stronger contributions.  The calculated AHC variation with energy suggests strong sensitivity to the chemical potential, highlighting the need for precise control over doping or external Fermi level shifts.  Such behavior is consistent with trends observed in other Heusler compounds, where the AHC is strongly dependent on the band structure and the degree of hybridization between transition metal d$-$states. For instance, in Co$_{2}$MnAl and Co$_{2}$MnSi, the AHC has been reported to be around 500 S cm$^{-1}$ and 1500 S cm$^{-1}$ \cite{Huang2015,Tung2013}, respectively, with the variation attributed to the different band dispersions near the Fermi level and the presence of gapped or nodal structures in the spin-polarized bands \cite{Noky2019}.  In comparison, the AHC in Mn$_{2}$CoAl, a compensated ferrimagnetic Heusler compound, reaches values as high as 1500-2000 S cm$^{-1}$ due to the presence of Weyl-like band crossings and enhanced Berry curvature contributions \cite{Kubler2012}. Similar trends have been observed in Fe-based Heusler compounds, such as Fe$_{2}$CrSi and Fe$_{2}$MnGa, where the interplay of band topology plays a crucial role in determining the magnitude of AHC \cite{Chadov2010}. Compared to these materials, Mn$_{2}$PdIn exhibits an intrinsic AHC due to the lower density of hybridized Mn$-d$, Pd$-d$, and In$-\{s, p\}$ states at the Fermi level. However, the significant increase in AHC when shifting the energy away from Fermi level suggests that this compound may host other extreme of anomalous AHC (very high) values under strain or doping induced modifications of the band structure. The observed energy dependence of AHC highlights the crucial role of band filling effects and suggests possible routes to enhance the AHC in similar compounds through controlled electronic structure engineering. The intriguing electronic structure and theoretical predictions motivate us to perform detailed transport and magneto-transport measurements on the sample.

As shown in \text{Fig.} \ref{F3}(a), zero-field $\rho_{xx}$ exhibits a metallic nature with a residual resistivity ratio ($\rho_{300 \mathrm{K}}/\rho_{2 \mathrm{K}}$) of 1.36, consistent with the characteristics of Heusler alloys \cite{li2020giant,bhattacharya2024spin,PhysRevLett.82.4280}. To investigate the principal scattering mechanism governing $\rho_{xx}$, we fit $\rho_{xx}(T)$ below $T_{f}$, employing the $e$-$e$ scattering $T^2$ dependence. However, a better fit is obtained on the incorporation of the spin-fluctuation $T^{3/2}$ term, resulting $\rho_{xx} = \rho_0 + aT^2 + bT^{3/2}$, where $\rho_0$ are residual resistivity and $a$, $b$ are constants \cite{mydosh1974low,PhysRevLett.82.4280}. The fitting yields $\rho_0 =$ 118.26 $\mu \Omega$ cm, $a=$ 1.28 $\times$ 10$^{-2}$ $\mu \Omega$ cm K$^{-3/2}$ and $b=$ 7.04 $\times$ 10$^{-4}$ $\mu \Omega$ cm K$^{-2}$, suggest a dominant spin-fluctuation contribution at low temperatures. This underscores a strong electron-magnon interaction between the localized moments and conduction electrons, consistent with the glassy nature of the magnetic ordering \cite{mydosh1974low}.

Now we focus on the Hall transport data. Figure \ref{F3}(c) depicts the $\rho_{xy}(H)$ isotherms at various $T$. Notably, alongside Lorentz force-mediated normal Hall resistivity $\rho_{xy}^O$, an additional component mirroring $M(H)$ appears in $\rho_{xy}$, in sharp contrast to parent SG Mn$_3$In\cite{chatterjee2020glassy}. This observation, coupled with the minuscule magnetoresistance [Fig. \ref{F3}(b)], suggest a dominant AHC contribution in $\rho_{xy}$. For a FM/FiM system, $\rho_{xy}$ is expressed as, $\rho_{xy} = \rho_{xy}^O + \rho_{xy}^A = R_0H + R_SM$, where $R_0$, $R_S$ are normal and anomalous Hall coefficients, respectively \cite{belopolski2019discovery,hurd2012hall,Ahmed2025b}. A linear fit of $\rho_{xy}$ in the high-field range ($H$ = 5 -7 T) yields $\rho_{xy}^A$ and the slope $R_0$, as shown in the inset of Fig. \ref{F3}(c). The negative slope of $\rho_{xy}(H)$ establishes electrons as the majority charge carriers, with a density of $\sim 10^{22}$ cm$^{-3}$ at $T$ = 5 K. Figure \ref{F3}(d) illustrates the non-monotonic variation of $\rho_{xy}^A$, obtained from high field extrapolation of $\rho_{xy}$ isotherms, showcasing a linear increase up to 150 K followed by a decrease with increasing temperature. Despite this temperature dependence, the AHC obtained by the tensorial relation $\sigma_{xy}^A \approx \rho_{xy}^A/\rho_{xx}^2$, exhibits a $T$ insensitive behaviour with $\sigma_{xy}^A \sim$ 50 S cm$^{-1}$ at $T$ = 5 K. Accounting in the various mechanisms governing AHC, $\rho_{xy}^A$ scales differently with $\rho_{xx}$ and the observed quadratic relation, in Fig. \ref{F3}(e), attests to the intrinsic ($int$) and/ or extrinsic side-jump ($sj$) rooted origin of AHC. Furthermore, the temperature independence of the anomalous Hall factor, $S_H = \sigma_{xy}^A/M$ [inset of Fig. \ref{F3}(d)], aligns well with the Karplus-Luttinger theory, reinforcing its validity in describing the observed AHC \cite{p13}.

To quantify the individual contributions, we adopt the \textit{TYJ}-scaling relations for $\rho_{xy}^A$, \textit{i.e.}, $\rho_{xy}^A/M\rho_{xx}$ = $a + b \rho_{xx}$. Here, $a$ is the skew-scattering coefficient and $b$ relates to the intrinsic AHC \cite{PhysRevLett.103.087206,RevModPhys.82.1539}. The expected linear relation (for $\rho_{xx}\leq 127$ $\mu\Omega$ cm), as in Fig. \ref{F3}(f), yields $a \approx$ -0.062 and $b \approx$ 63 S cm$^{-1}$. The deviation from linearity for $T>$ 120 K originates from the temperature dependence of $M$. Using these coefficients, we calculated the skew scattering term (a$\rho_{xx}$) and the combined $int$ and $sj$ contribution (b$\rho_{xx}^2$), plotting their temperature variations [inset of Fig. \ref{F3}(f)]. This reveals that the contributions of $int$ and $sj$ mechanisms dominate the AHC across the investigated temperature range. At low temperatures, the reduced phonon scattering entangles $\sigma_{sj}^A$ and $\sigma_{xy, int}^A$, present a challenge for quantification without an existing scaling framework. However, the estimated order of magnitude of $\sigma_{sj}^A$ is significantly smaller than $\sigma_{xy, int}^A$, affirming the intrinsic nature of AHC in Mn$_2$PdIn.

 We found that the experimentally obtained AHC for the sample is lower than that  DFT calculated AHC (0 K). This discrepancy can likely be attributed to the polycrystalline nature of the sample, defects and/or finite-temperature effects, which are known to significantly reduce AHC by 50-80\% compared to theoretical predictions \cite{Kubler2012,Kudrnovsky2013}.  The calculated AHC variation with energy suggests strong sensitivity to the chemical potential, highlighting the need for precise control over doping or external Fermi level shifts.

In conclusion, Mn$_2$PdIn, an inverse Heusler alloy, exhibits two distinct spin-glassy phases within a nearly compensated ferrimagnetic background, along with a notably large intrinsic anomalous Hall conductivity. These characteristics position it as a promising candidate for spintronic applications, offering both quenched magnetization and robust topological transport properties. The quadratic scaling of anomalous Hall resistivity with longitudinal resistivity confirms the intrinsic origin of the anomalous Hall conductivity in this compound. Ab initio calculations reveal Weyl-type band crossings near the Fermi level, attributed to the breaking of time-reversal ($\mathcal{T}$) symmetry, thereby validating its topologically non-trivial electronic structure.  We propose that the anomalous Hall effect in Mn$_2$PdIn is deeply rooted in the material's band topology and pronounced Fermi surface nesting. Our findings present a predictive criterion-nesting-induced inter-orbital scattering in spin-orbit-coupled systems-for identifying new anomalous Hall effect-active intermetallic compounds with Heusler and related crystal structures.

Works, performed  at the Ames National Laboratory was supported by the Division of Materials Science and Engineering of the Office of Basic Energy Sciences, Office of Science of the U.S. Department of Energy (D.O.E). Ames National Laboratory is operated for the U.S. DOE by Iowa State University of Science and Technology under Contract No. DE-AC02-07CH11358. A.A. and A.B. would like to acknowledge SINP, India, and the Department of Atomic Energy (DAE), India for their fellowship. We acknowledge Mohammad Rezwan Habib and Prof. Indra Dasgupta from the School of Physical Sciences, Indian Association for the Cultivation of Science, Kolkata, for the initial theoretical assessment and fruitful discussions. 

\end{document}